\documentclass[aps,pre,superscriptaddress,floatfix,10pt]{revtex4}
\usepackage{graphicx}
\usepackage{dcolumn}
\usepackage{bm}
\usepackage{latexsym}
\usepackage{amsmath}
\usepackage{amssymb}
\usepackage{color}

\begin{document}

\title{First Passage Time in Computation by Tape-Copying Turing Machines:\\ Slippage of Nascent Tape}
 \author{Soumendu Ghosh}
\affiliation{Department of Physics, Indian Institute of Technology
  Kanpur, 208016}\author{Shubhadeep Patra} \affiliation{ISERC, Visva-Bharati, Santiniketan 731235}
\author{Debashish Chowdhury{\footnote{Corresponding author; e-mail: debch@iitk.ac.in}}}
\affiliation{Department of Physics, Indian Institute of Technology
  Kanpur, 208016}
  
\begin{abstract}
Transcription of the genetic message encoded chemically in the sequence of the DNA template is carried out by a molecular machine called RNA polymerase (RNAP). Backward or forward slippage of the nascent RNA with respect to the DNA template strand give rise to a transcript that is, respectively, longer or shorter than the corresponding template. We model a RNAP as a ``Tape-copying Turing machine'' (TCTM) where the DNA template is the input tape while the nascent RNA strand is the output tape. Although the TCTM always steps forward the process is assumed to be stochastic that has a probability of occurrence per unit time. The time taken by a TCTM for each single successful forward stepping on the input tape, during which the output tape suffers lengthening or shortening by $n$ units because of backward or forward slippage, is a random variable; we report some of the statistical characteristics of this time by using the formalism for calculation of the distributions of {\it first-passage time}. The results are likely to find applications in the analysis of experimental data on ``programmed'' transcriptional error caused by transcriptional slippage which is a mode of ``recoding'' of genetic information.
\end{abstract}

\maketitle
\section{Introduction}

Time taken by a system to reach one specific state {\it for the first time} starting from another specified initial state is defined as the corresponding {\it first-passage time } (FPT). In case of stochastic processes \cite{bressloff14} FPT is a random variable and the distribution of First-passage times (DFPT) is used for quantitative statistical characterization of the transition. DFPT has been calculated for wide varieties of physical processes in nonliving as well as in living systems 
\cite{redner01,redner14,chou14,biswas16,godec16}. Markov processes \cite{gillespie92}, which are memoryless stochastic processes, are often good approximations for real stochastic phenomena.  For mathematical formulation of these processes one defines the probabilities of all the discrete states of the system and prescribes the rates for all the allowed transitions between these states; the time evolution of the probabilities are described mathematically in terms of master equations. 

Operations of almost all types of molecular machines and devices have been formulated in terms of master equations \cite{chowdhury13a,chowdhury13b,kolomeisky15}. DFPT calculated from the master equations for the kinetics of these machines often correspond to physical quantities that are experimentally measurable. For example, the duration of dwell of a molecular motor at a given position on its track is a FPT. The dwell time distribution of various types of molecular motors have received attention in the past \cite{kolomeisky05,bierbaum13,tripathi09,sharma11,sharma12,sharma13,thompson15}. In this paper we calculate DFPT for a class of stochastic processes in a simplified theoretical model of operation of a molecular machine called RNA polymerase (RNAP) \cite{buc09}.  The biologically motivated, but highly simplified, model of molecular machine that we analyze here can be interpreted as a physical realization of a Turing machine \cite{feynman} which is an idealized device conceptualized for abstract `computation' \cite{minsky67}. 

In the simplest formulation, a finite set of discrete states are assigned to the machine. It can move forward and backward on a tape in discrete steps; the step size being the size of the boxes marked on the tape. Each box on the tape stores a digit. The head of the Turing machine can read the digit stored in the box at its current location. The Turing machine reads this `input' and the result of its `computation' is an `output' digit and a concomitant transition of state of the machine according to the fixed set of rules (algorithm) prescribed in the beginning. 

A single DNA strand serves as the tape for a RNAP and the sequence of nucleotides are the analogues of sequence of digits stored on the tape. The biochemical and conformational states of the RNAP are the counterparts of the internal states of a Turing machine. However, in contrast to output digits of a Turing machine the output of the `computation' (which biologists refer to as transcription) by the RNAP  is another tape called RNA. Thus, a RNAP is a `tape-copying Turing machine' (TCTM) \cite{bennett82,mooney98,sharma12b}. In order to adopt unambiguous terminology we refer to the tape on which input data are engraved as the `input' tape while the incomplete output tape during ongoing elongation is also referred to as the `nascent' tape. 

In our model the TCTM ia assumed to move always forward by a single step on the input tape. But its forward stepping is assumed to be a stochastic process with a probability of occurrence per unit time. Each forward stepping of the TCTM completes one step of the computation. Completion of $L$ such steps of computation by the RNAP in a perfect error-free manner on an input DNA template tape of length $L$ would result in an output RNA tape whose length (both in the units of nucleotides) is also $L$, i.e., exactly equal to that of the input DNA tape. But, as we explain in the next section, a shorter/longer RNA tape would result from an erroneous forward/backward `slippage' of the nascent output tape at any step of the sequence of computations by the RNAP \cite{anikin10}. One of the fundamental questions is the dwell time of a RNAP at a nucleotide on the template DNA during which the nascent transcript associated with it suffers slippages causing it to become longer (or shorter) by $n$ nucleotides. In the terminology of computation, this is equivalent to the dwell time of the reading tip of the TCTP at a single position on the input tape during which the nascent output tape becomes longer (or shorter) by $n$ units because of its slippage. This time is intrinsically stochastic and can be formulated as a first-passage time; we calculate its statistical distribution in this paper.

In the past, the dwell time distribution, which is a DFPT, has been calculated for simple models of RNAP \cite{tripathi09}. However, those distributions characterize the stepping pattern of the RNAP, i.e., the statistics of the time taken by the Turing machine to perform each step of computation that produces an error-free output. In contrast, in this paper, we focus on the erroneous process of `slippage' of the output RNA tape and calculate a new DFPT that characterize the statistical features of this slippage process. More precisely, the type of slippage we consider here take place only in so-called elongation stage of transcription when the nascent RNA tape gets elongated in each round of elementary computation by the RNAP machine.

\section{Biological motivation: slippage of nascent RNA}

\begin{figure}[t]
(a)\\
\includegraphics[angle=0,width=0.45\columnwidth]{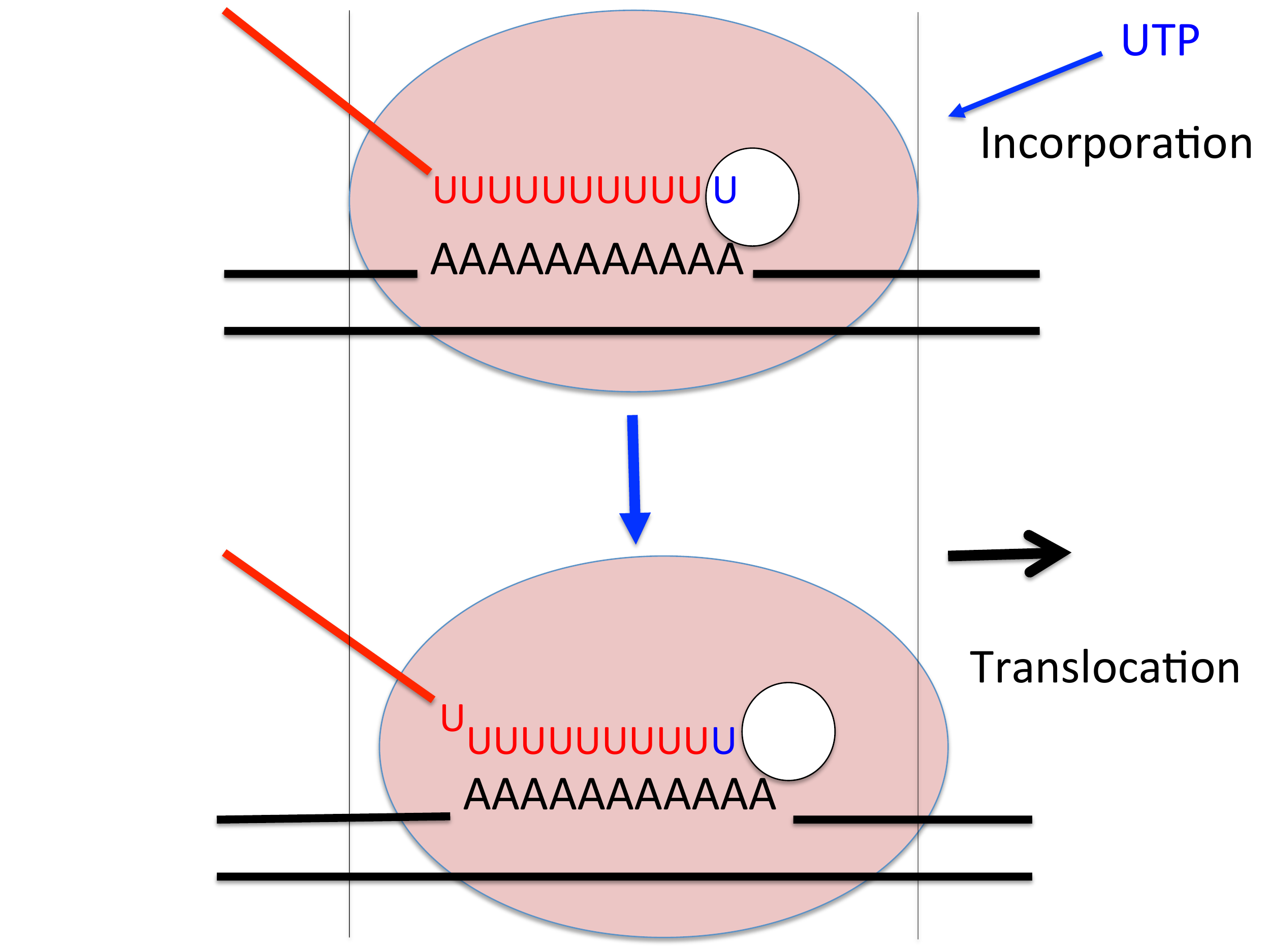}\\[0.05cm]
(b)~~~~~~~~~~~~~~~~~~~~~~~~~~~~~~~~~~~~~~~~~~~~~~~~~~~~~~~~~~~~~~~~~~~~~~~~~~(c)\\
\includegraphics[angle=0,width=0.45\columnwidth]{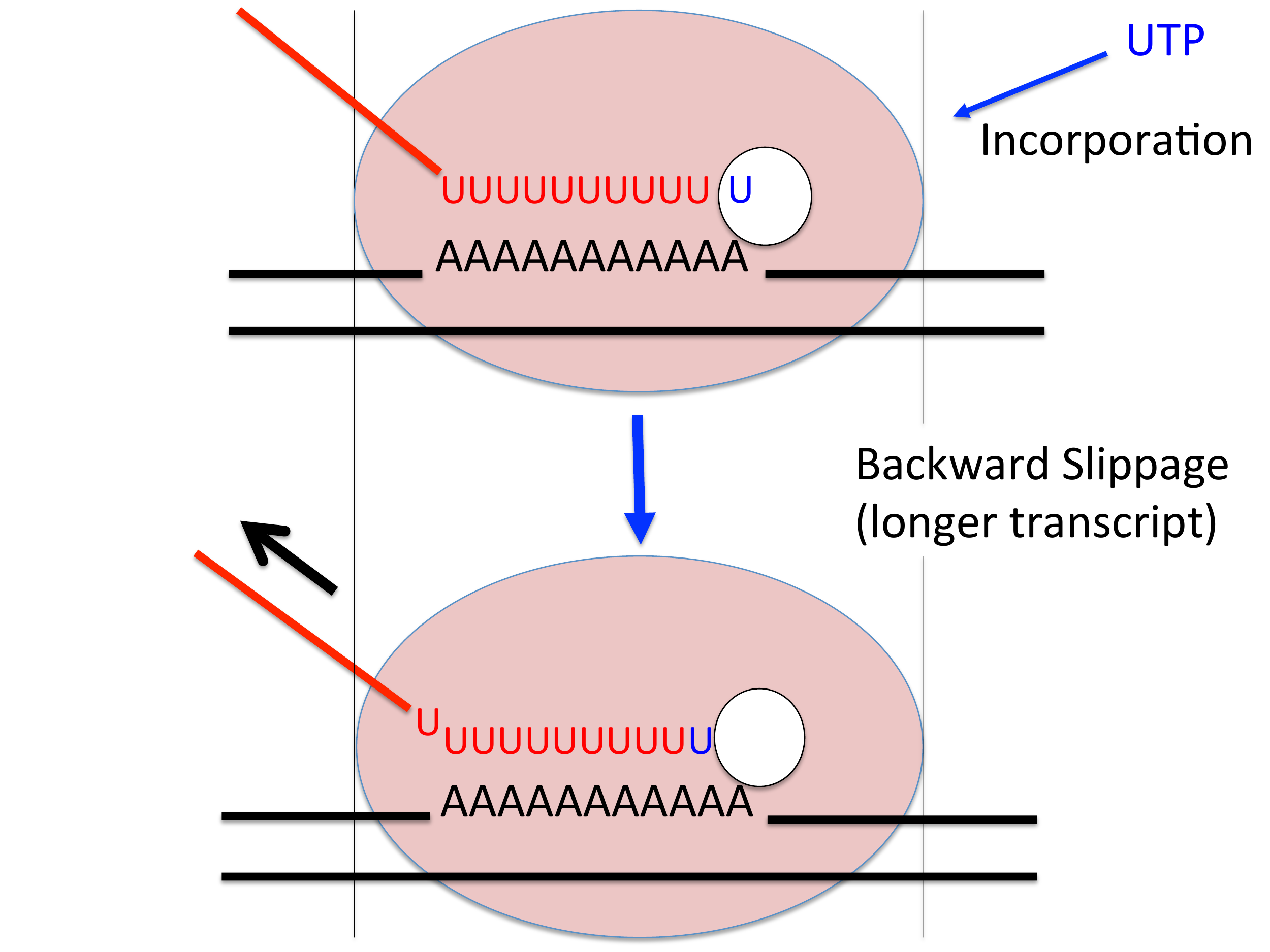}
\includegraphics[angle=0,width=0.45\columnwidth]{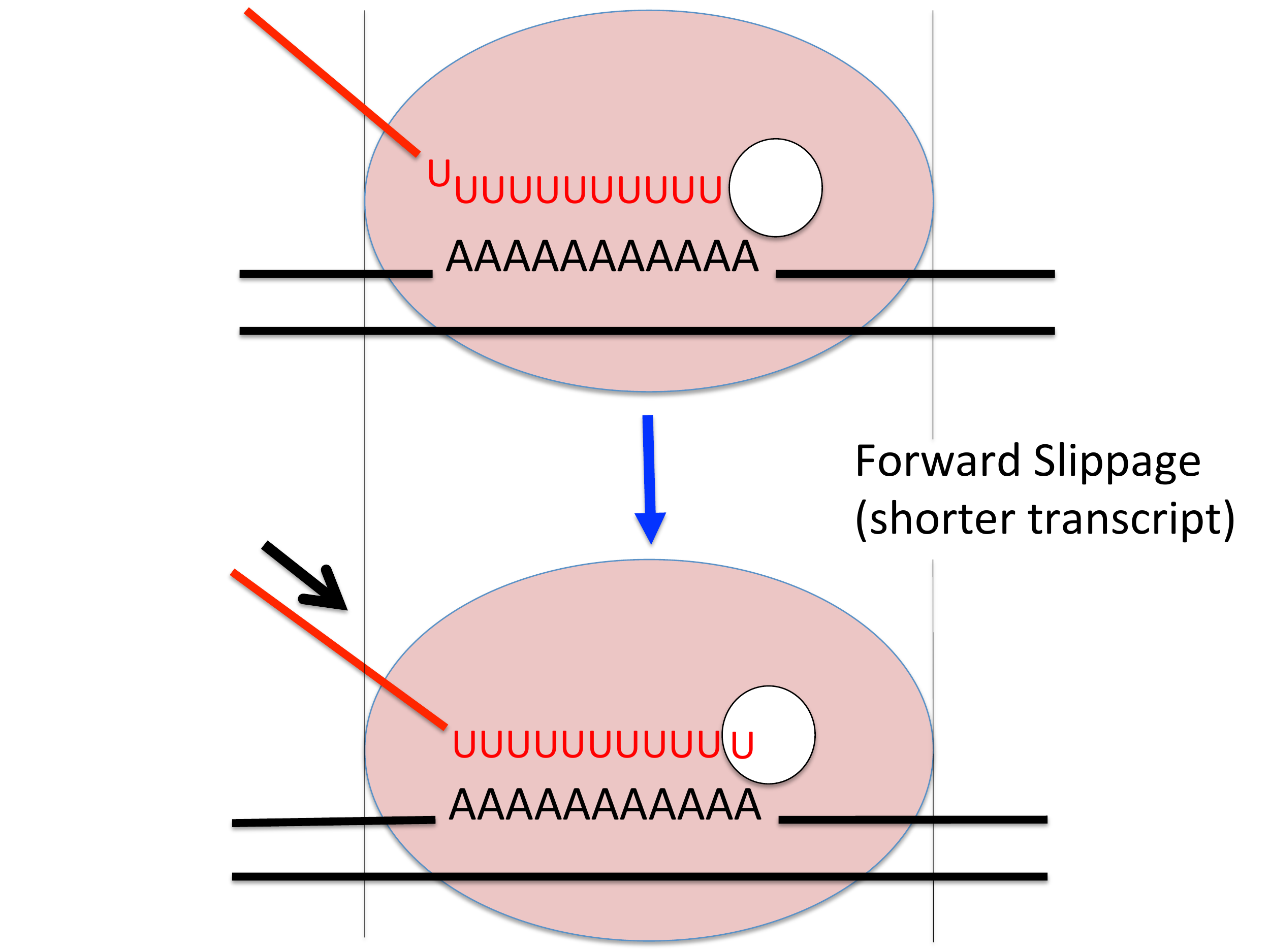}\\[0.05cm]
\caption{A schematic diagram illustrating (a) error-free transcription, without slippage, by a RNAP, (b) backward slippage of nascent RNA, (c) forward slippage of nascent RNA. }
\label{fig-slippage_model}
\end{figure}

The phenomenon of slippage of the nascent RNA during transcription (i.e., its synthesis by a RNAP machine), keeping the position of the machine fixed on the template DNA is depicted schematically in Fig.\ref{fig-slippage_model}. 

\begin{figure}[h]
\begin{center}
\includegraphics[angle=0,width=0.5\columnwidth]{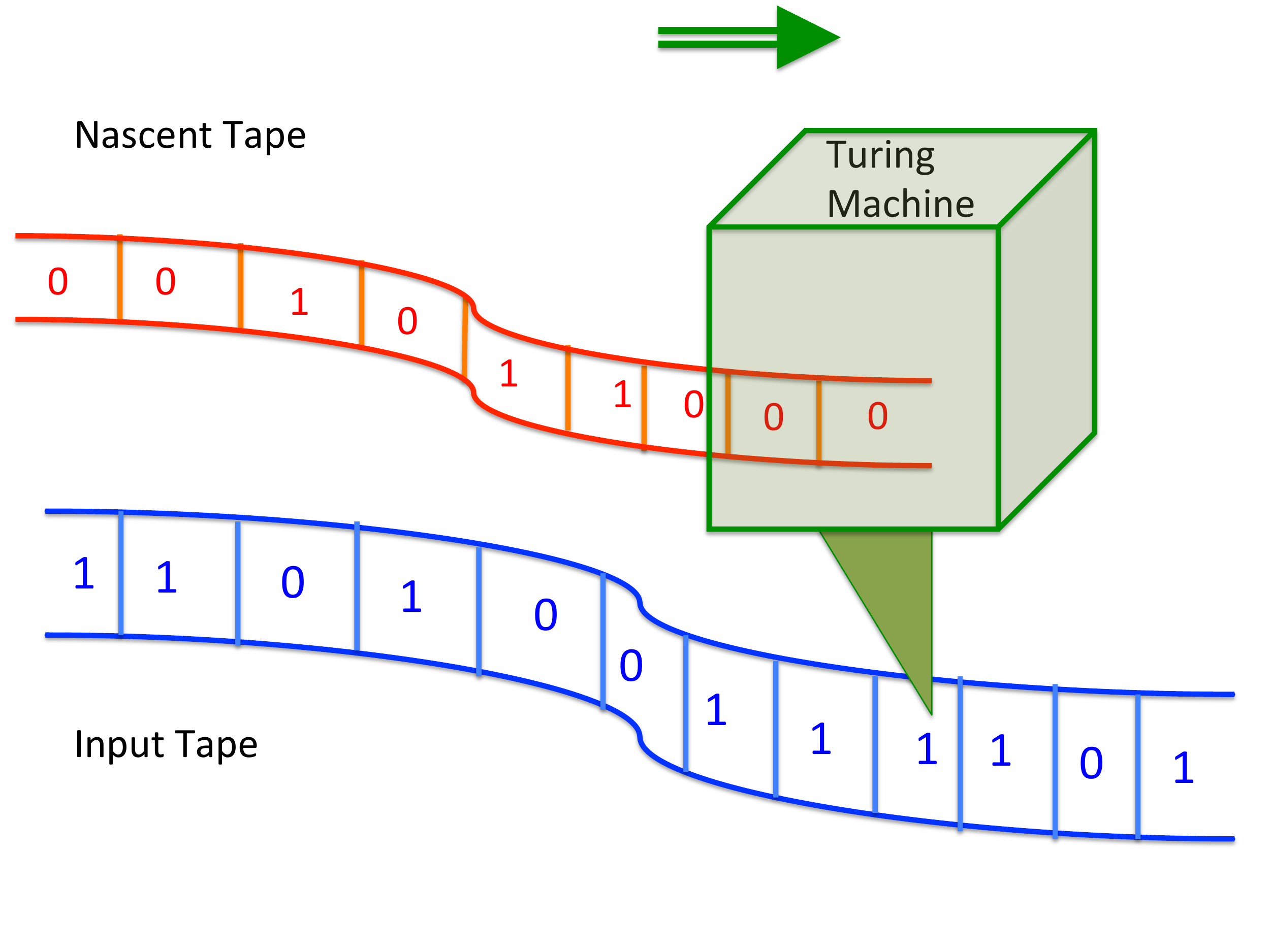}\\[0.02cm]
\end{center}
\caption{A schematic diagram of the tape copying Turing machine.}
\label{fig-Turing_Machine}
\end{figure}

\begin{figure}[h]
(a)\\
\includegraphics[angle=0,width=0.45\columnwidth]{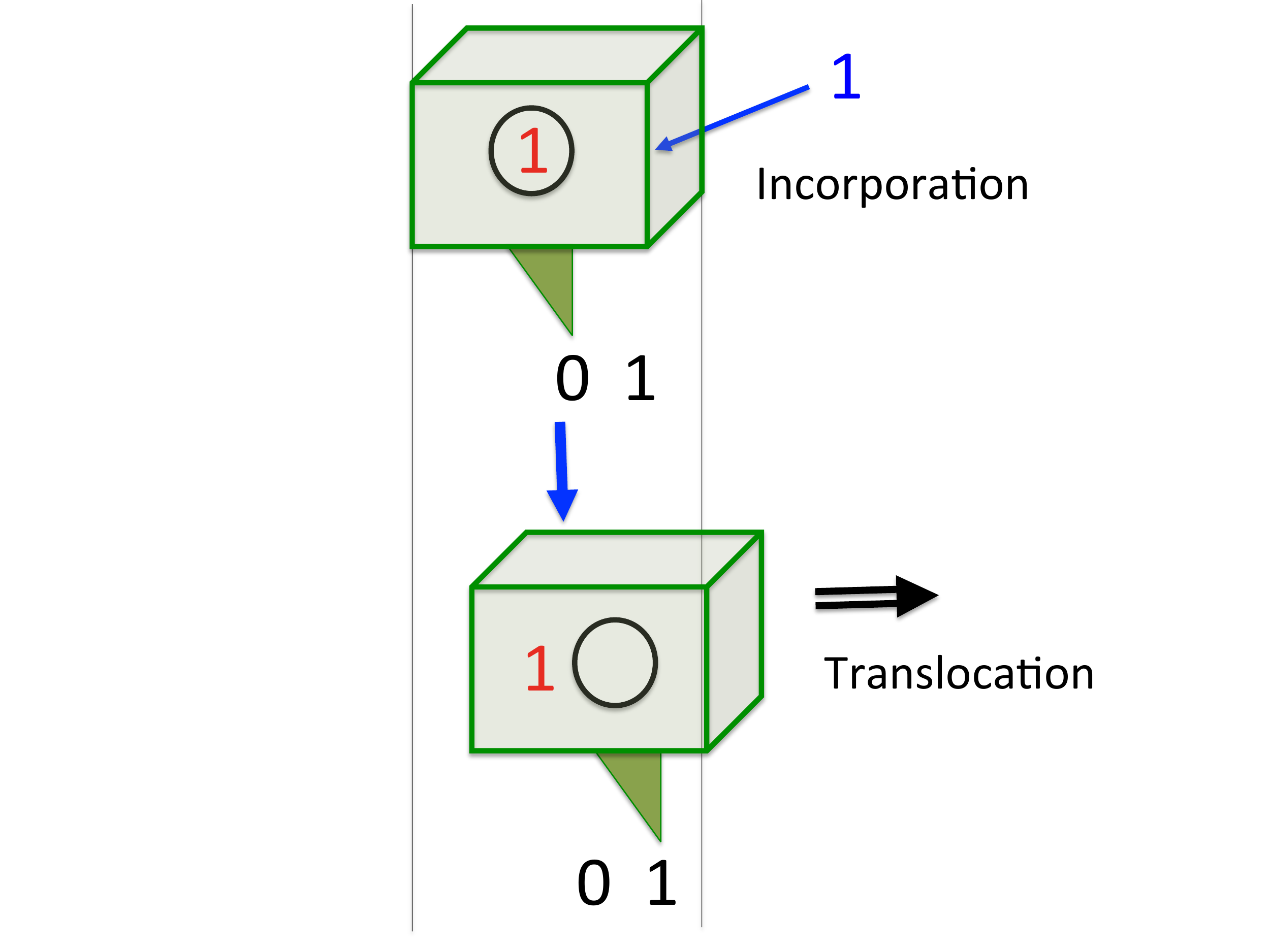}\\
(b)~~~~~~~~~~~~~~~~~~~~~~~~~~~~~~~~(c)\\
\includegraphics[angle=0,width=0.45\columnwidth]{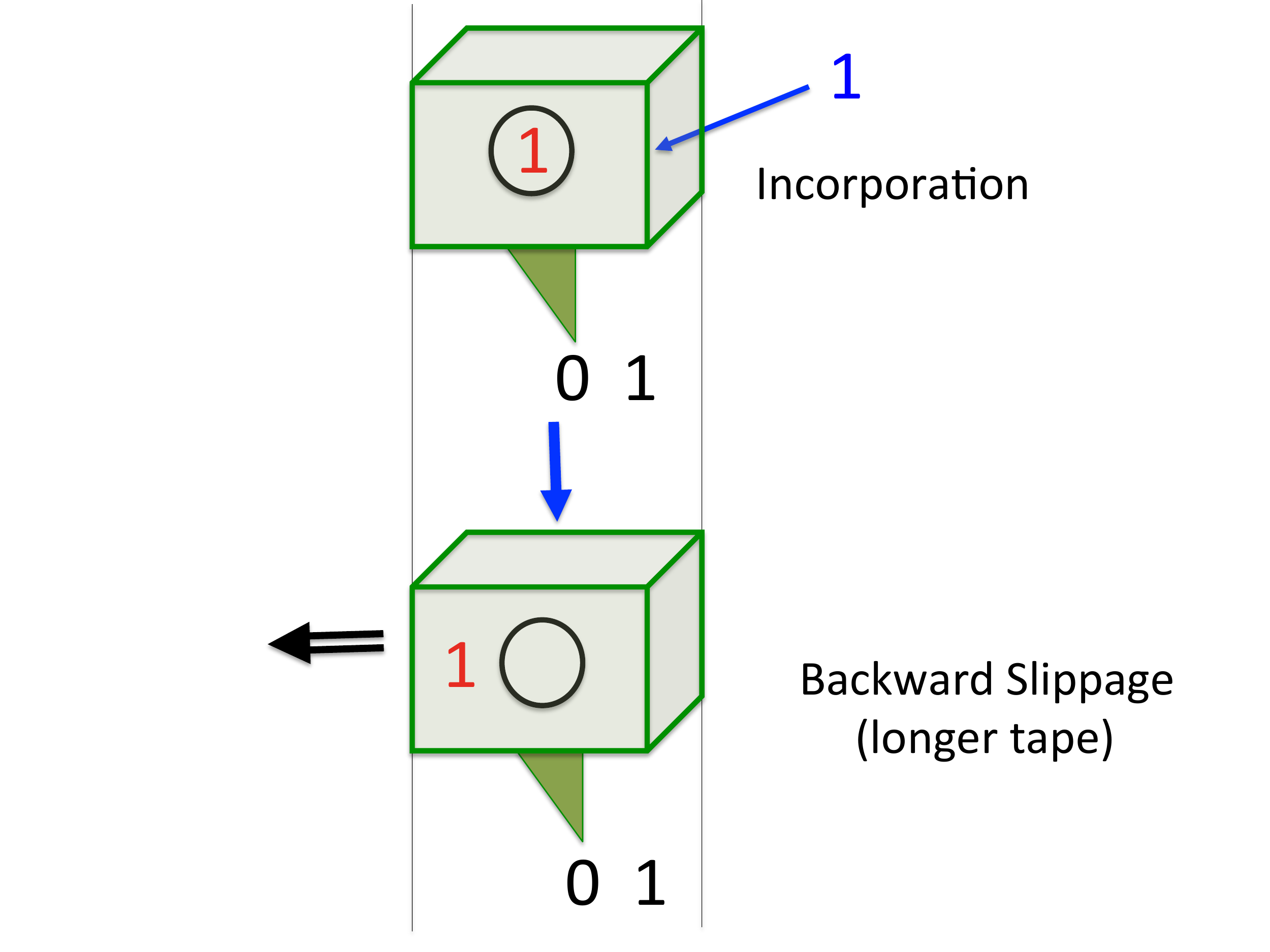}
\includegraphics[angle=0,width=0.45\columnwidth]{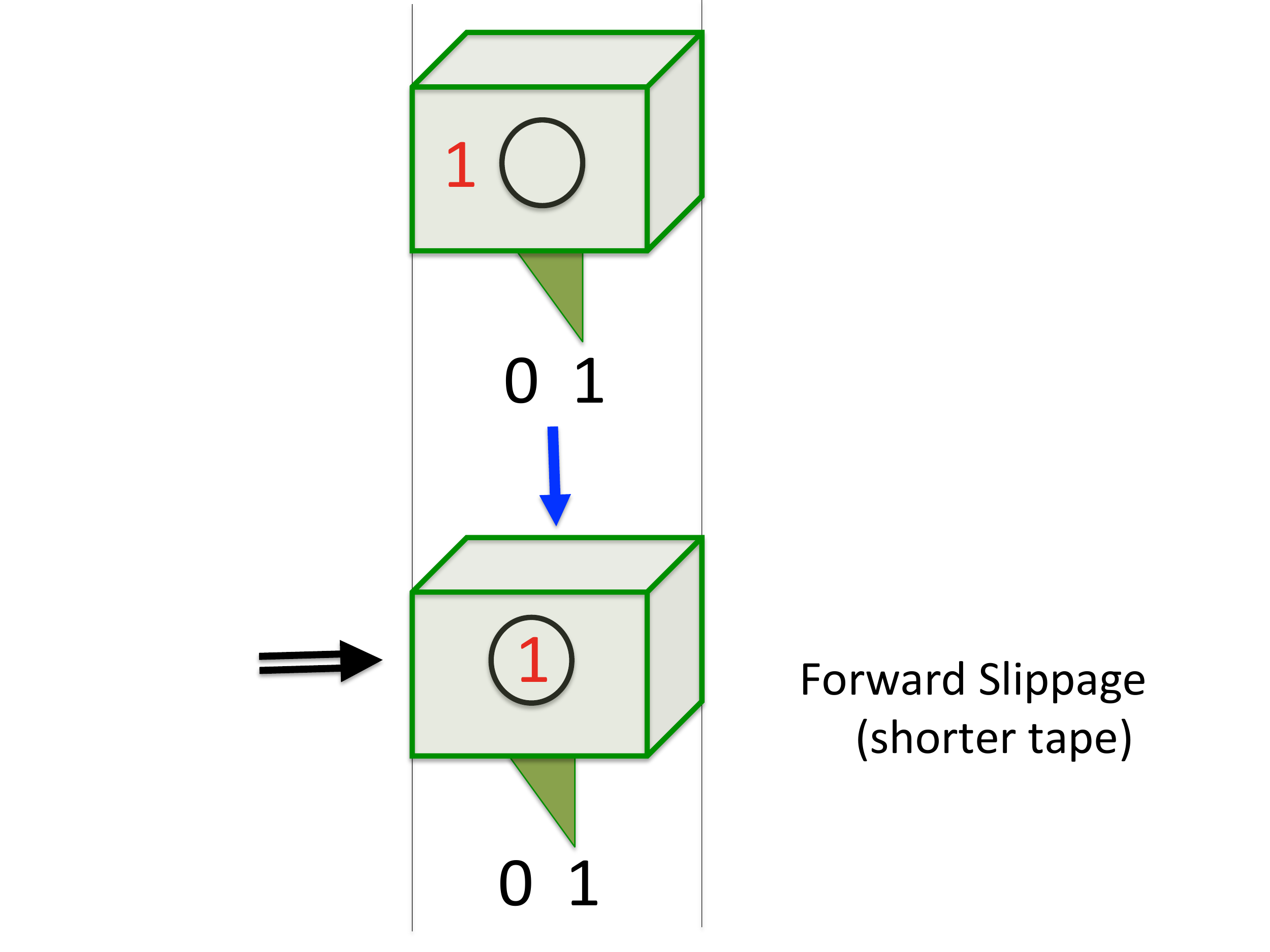}
\caption{A schematic diagram of transcript slippage. Circle represents the head of the Turing machine. (a) translocation, (b) backward slippage and (c) forward slippage of Turing machine.}
\label{fig-transcript_slippage_model}
\end{figure}

In the four-letter alphabet used for encoding genetic message, the letter `A' on the DNA template is complementary to the letter `U' on the RNA transcript. The grip of the RNAP on the template DNA as well as on the nascent RNA transcript are important for error-free normal transcription. Loosening of its grip on the template DNA can lead to its backward - or forward-slippage along the template DNA; the cause of this phenomenon has received lot of attention over the last decade \cite{zhang16}. In contrast, in this paper we focus on the consequence of lose grip of the RNAP on the nascent transcript that can cause backward or forward slippage of the transcript (see Fig.\ref{fig-slippage_model}) while the position of the RNAP on the template DNA remains unaltered. Such transcriptional errors are believed to be `programmed', rather than random, and constitute one specific mode of `recoding' of the genetic message \cite{atkins10}.

\begin{figure}[h]
\begin{center}
\includegraphics[angle=0,width=0.5\columnwidth]{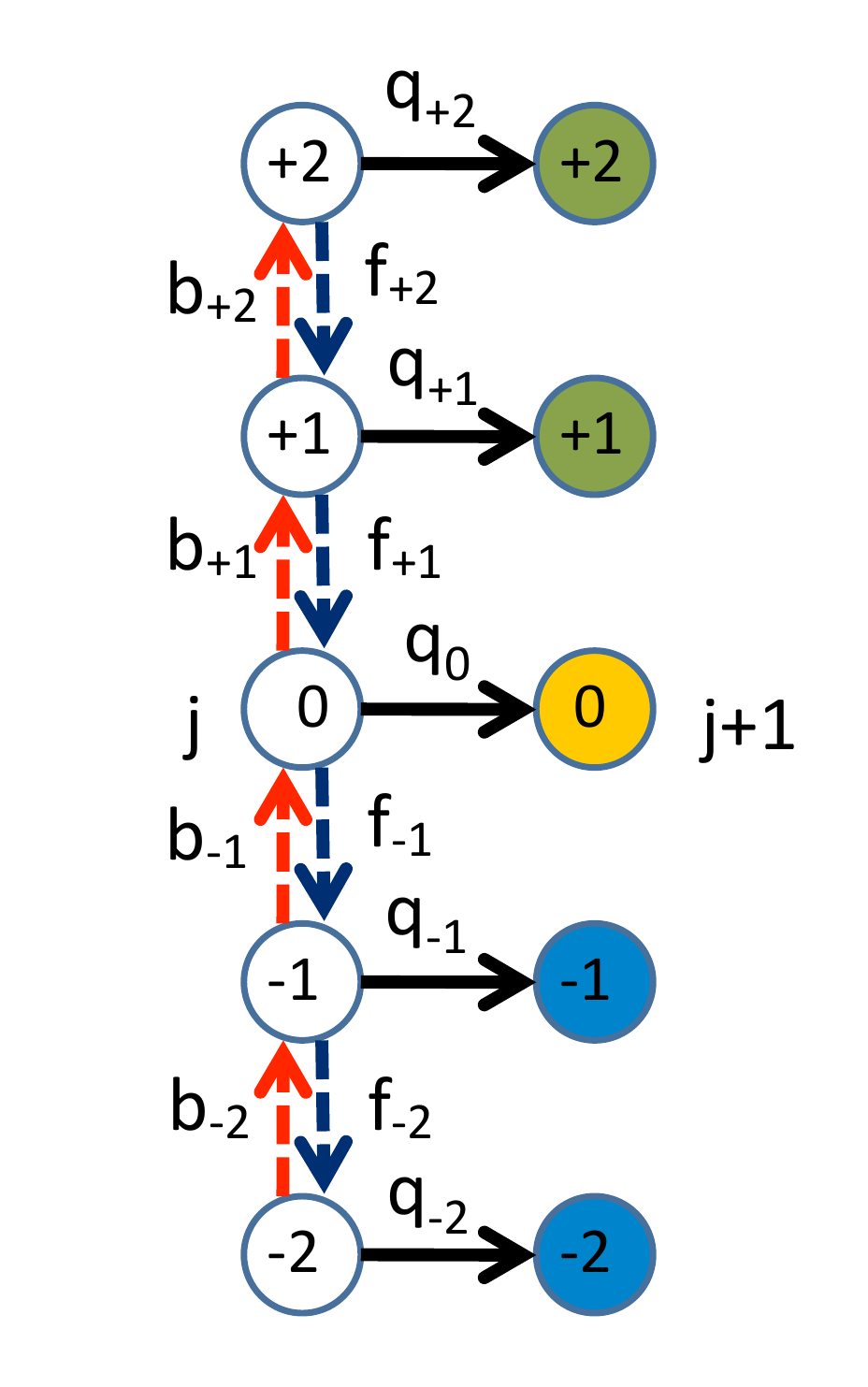}\\[0.02cm]
\end{center}
\caption{A schematic diagram of the kinetic model for the special case where the maximum allowed value of $|\mu|$ is $2$ (referred to as the 5-state model).  b's and f's are the rates of backward and forward slippages, respectively. q's are the rates of translocation of the TM from j-th to j+1-th position on the input tape.}
\label{fig-FP_5_state_model}
\end{figure}

\section{Theoretical model: slippage of nascent tape}

As stated in the introduction, this model is motivated by the biological phenomenon of RNA slippage during transcription which has been explained briefly in the preceding section. The input tape and the nascent tape in our theoretical model mimic the DNA template and nascent RNA respectively, while the tape-copying Turing machine represents the RNAP.  We assume that the input tape passes through a channel inside the reader of the Turing machine. At any given instant of time, the head of the reader (a tip) can read only one of the boxes on the input tape in the segment located inside the channel. Similarly, a segment of the nascent tape, starting from its elongating tip, also remains in the same channel inside the reader. As the nascent tape elongates adding one box at each step of computation it emerges from the other end of the channel. We assume that the channel inside the reader covers several boxes on the input and nascent tapes  at a time and that this channel constitutes a slippage prone tract. 

The slippage is known to occur on a homogeneous sequence of identical nucleotides on the template DNA although on a longer scale the sequence on the DNA is inhomogeneous. Since in this paper we focus exclusively on the slippage process we assume a homogeneous sequence on the input tape. For the sake of simplicity we denote the homogeneous sequence of digits on the  template tape by a string of 0 (zero). We implement the complementary base-pairing between the template DNA and nascent RNA by the programming the Turing machine to perform the logical operation XOR (exclusive OR) which gives output 1 (one) whenever the input is 0 (zero) and vice-versa. 

The instantaneous position of the head of the reader of the TM is denoted by the integer index $j$; it takes a forward step from $j$ to $j+1$ after a successful error-free step of computation. The extra length of the nascent tape caused by the slippage is labelled by an integer index $\mu$ that can, in principle, be positive, negative or zero; $\mu =0$ if the nascent tape suffers no slippage or it suffers equal numbers of forward and backward slippages. In contrast, $\mu$ is positive (negative) in case backward (forward) slippage occurs more often than the reverse process (see Fig.\ref{fig-FP_5_state_model} where the model has been illustrated for the maximum value $|\mu|=2$). In other words, $\mu$ is the excess or deficiency of boxes (``extra'' number of boxes) in the nascent tape as compared to the corresponding template. From now onwards, a TM associated with a nascent tape of ``extra'' length $\mu$ will be referred to as a TM in slippage state $\mu$; a negative value of $\mu$ indicates shorter length of the nascent tape as compared to the template tape. The instantaneous state of the TM will be denoted by the pair $j,\mu$.

\subsection{First-Passage Time Distributions and first two moments} 

The time taken by the TM to reach, {\it for the first time}, the state $j+1,\mu$,  from the state $j,0$ is defined as the {\it first-passage time} ${\cal P}_{\mu}(t)$ for the ``slippage state'' $\mu$; since all the sites on the template are equivalent the site index is not explicitly written in the symbol for the first-passage time. Suppose, for the $n$-state model, $P_{\mu}^{(n)} (j,t)$ denotes the probability, at time $t$, that the head of the TM is located at the $j$-th site on its template tape and is in the slippage state $\mu$.  

In principle, if the full distribution ${\cal P}_{\mu}^{(n)}(t)$ of the first-passage times is known the corresponding 
mean first-passage times $< t_{\mu} >$ can be computed from the definition
\begin{eqnarray}
< t_{\mu}^{(n)} >&=&\int_{0}^{\infty} t ~{\cal P}_{\mu}^{(n)}(t)~ dt
\end{eqnarray}
while the randomness parameter can be obtained using the definition
\begin{eqnarray}
r_{\mu}^{(n)}&=&\frac{ < (t_{\mu}^{(n)})^{2} > - < t_{\mu}^{(n)} >^{2}}{ < t_{\mu}^{(n)} >^{2}}
\end{eqnarray}
where 
\begin{eqnarray}
<(t_{\mu}^{(n)})^{2} >&=&\int_{0}^{\infty} t^{2} ~{\cal P}_{\mu}^{(n)}(t)~ dt
\end{eqnarray}
However, most often the set of kinetic equations are too complicated to yield closed form analytical expression for ${\cal P}_{\mu}^{(n)}(t)$. In such situations the moments of the distribution of first passage time can still be obtained by taking appropriate derivatives of $P_{\mu}^{(n)}(s)$ if the latter can be calculated in the s-space (Laplace space). For example,
the normalised mean first-passage time can be obtained using 
\begin{eqnarray}
<t_{\mu}^{(n)}>&=&\frac{-\frac{d}{ds} P_{\mu}^{(n)} (s) \Big|_{s=0}}{P_{\mu}^{(n)} (s) \Big|_{s=0}}
\label{eq-meanDERlaplace}
\end{eqnarray}

For the sake of simplicity, we write the master equations only for $\mu=0,\pm 1, \pm 2$ which correspond to the 5-state kinetic model in Fig. \ref{fig-FP_5_state_model}. In this case the master equations governing the time evolution of $P_{\mu} (j,t)$ 
can be written as a matrix differential equation. 
So, in compact form (for simplicity we omit the j indices)
\begin{eqnarray}
\frac{d\textbf{P}(t)}{dt}&=&A~ \textbf{P}(t).
\label{matrix-eq1}
\end{eqnarray}
where
\begin{equation}
\textbf{P}(t) = {\begin{pmatrix}~ P_{0}(j,t)~ \\ \\ P_{+1}(j,t) \\ \\ P_{+2}(j,t) \\ \\ P_{-1}(j,t) \\ \\~ P_{-2}(j,t)~\end{pmatrix}}
\end{equation}
and 
\begin{widetext}
\begin{eqnarray}
A = 
{\begin{pmatrix}~ 
-(q_{0}+b_{+1}+f_{-1}) & f_{+1} & 0 & b_{-1} & 0 \\ \\ b_{+1} & -(q_{+1}+b_{+2}+f_{+1}) & f_{+2} & 0 & 0 \\ \\ 0 & b_{+2} & -(q_{+2}+f_{+2}) & 0 & 0 \\ \\ f_{-1}  & 0 & 0 & -(q_{-1}+b_{-1}+f_{-2}) & b_{-2} \\ \\ 0 & 0 & 0 & f_{-2} & -(q_{-2}+b_{-2})
~\end{pmatrix}}  \nonumber \\
\label{matrix-eqn} 
\end{eqnarray}
\end{widetext}

For the calculation of the first-passage time, we impose the initial conditions 
$P_{\mu}(0) = 0$ for all $\mu$ except $P_{0}=1$. Carrying out Laplace transform 
\begin{eqnarray}
\mathcal{L}\biggl[\frac{d\textbf{P}(t)}{dt}\biggr] = A ~\mathcal{L} ~\textbf{P}(t) 
\end{eqnarray}
we get
\begin{eqnarray}
s \tilde{\textbf{P}}(s) - \textbf{P}(0) = A~\tilde{\textbf{P}}(s) 
\end{eqnarray}
and hence
\begin{eqnarray}
\tilde{\textbf{P}}(s)&=&(s \textbf{I} - A) ^{-1}~\textbf{P}(0)
\label{laplace-eq}
\end{eqnarray}
where $\textbf{I}$ is the identity matrix, $\mathcal{L}$ indicates the Laplace transform  operator and $\tilde{\textbf{P}}(s)$ is the Laplace transform of $\textbf{P}(t)$, i.e., $\tilde{\textbf{P}}_{\mu}(s)$ is the Laplace transform of $\textbf{P}_{\mu}(t)$.
After taking inverse Laplace transform of $\tilde{\textbf{P}}(s)$, in principle, one would get
\begin{eqnarray}
\textbf{P}(t)&=&\mathcal{L}^{-1}~ [ ~\tilde{\textbf{P}}(s) ~]
\end{eqnarray}

\section{Results}

For the sake of simplicity of graphical plots, we present the results here only for the cases, namely $b_{+1}=b_{+2}=b_{-1}=b_{-2}=b$ , $f_{+1}=f_{+2}=f_{-1}=f_{-2}=f$ and $q_{0}=q_{+1}=q_{+2}=q_{-1}=q_{-2}=q$. Moreover, from now onwards, the special case $b=f$ will be referred to as the {\it symmetric} case while the more general situation $b\neq f$ will be referred to as the {asymmetric case}. The analytical expressions that we derive for the 3-state model are given in the main text of this section because these are not only short but also display some systematic trends which can be exploited for physical interpretation. The analytical expressions of the results for 5-state and 7-state models are simple enough to be reproduced in the main text. For all the graphical plots we have taken $q=10~s^{-1}$. In each of the figures we draw the curves obtained from the theoretical expressions.

\subsection{Results for the general n-state model in the symmetric case}

We get the generalised form of the probabilities in Laplace space,
\begin{eqnarray}
P_{\mu}^{(n)} (s)&=& \frac{\underbrace{b^{\mu}(b+q+s)^{\lambda-\mu}}_{{\rm for}~ \lambda \ge \mu}+ \underbrace{(\lambda-\mu-1) b^{\mu+1}(q+s)^{\lambda-\mu-1}}_{{\rm for}~ \lambda - \mu \ge 2}+ \underbrace{3 {\lambda-\mu-1 \choose \lambda-\mu-3}b^{\mu+2}(q+s)^{\lambda-\mu-2}}_{{\rm for}~ \lambda - \mu \ge 3}+{\rm higher ~order~ terms}}{(q+s)\biggl\{ \underbrace{n b^{\lambda}+(q+s)^{\lambda}}_{{\rm for} \lambda \ge 1}+\underbrace{n b (q+s)^{\lambda-1}}_{{\rm for} \lambda \ge 2}+\underbrace{(\lambda-1) n b^{2} (q+s)^{\lambda-2}}_{{\rm for} \lambda \ge 3}+{\rm higher ~order~ terms}\biggr\}}\nonumber \\
\end{eqnarray}
where $n$ is the total number of states, $\mu$ denotes the slippage state and 
$\lambda=\mu_{max}$ is the maximum possible slippage state for a given n, i.e., 
\begin{eqnarray}
\mu \le \lambda = \mu_{max} = (n-1)/2.
\end{eqnarray}

For example,
\begin{eqnarray}
\frac{<t_{\mu}^{(3)}>}{<t_{0}^{(3)}>}&=&\frac{n \Bigl(\frac{b}{q}\Bigr)^{2}+(n\mu+2)\Bigl(\frac{b}{q}\Bigr)+\mu+1}{n\Bigl(\frac{b}{q}\Bigr)^{2}+2\Bigl(\frac{b}{q}\Bigr)+1}\\ \nonumber
\\
\frac{<t_{\mu}^{(5)}>}{<t_{0}^{(5)}>}&=&\frac{n\Bigl(\frac{b}{q}\Bigr)^{5}+n(\mu+4)\Bigl(\frac{b}{q}\Bigr)^{4}+(4n\mu+4n+3)\Bigl(\frac{b}{q}\Bigr)^{3}+(4n\mu+n+\mu+10)\Bigl(\frac{b}{q}\Bigr)^{2}+(n\mu+3\mu+6)\Bigl(\frac{b}{q}\Bigr)+\mu+1}{n\Bigl(\frac{b}{q}\Bigr)^{5}+3n\Bigl(\frac{b}{q}\Bigr)^{4}+(4n+3)\Bigl(\frac{b}{q}\Bigr)^{3}+(2n+9)\Bigl(\frac{b}{q}\Bigr)^{2}+7\Bigl(\frac{b}{q}\Bigr)+1}\nonumber\\
\\
\frac{<t_{\mu < \lambda}^{(7)}>}{<t_{0}^{(7)}>}&=&\frac{term1-term2}{term3}
\end{eqnarray}
where
\begin{eqnarray}
term1&=&\biggl\{n\Bigl(\frac{b}{q}\Bigr)^{6}+10n\Bigl(\frac{b}{q}\Bigr)^{5}+32n\Bigl(\frac{b}{q}\Bigr)^{4}+(39n+4)\Bigl(\frac{b}{q}\Bigr)^{3}+(19n+24)\Bigl(\frac{b}{q}\Bigr)^{2}+(3n+20)\Bigl(\frac{b}{q}\Bigr)+4 \biggr\} \times \nonumber \\
&& \Bigl\{\Bigl(\frac{b}{q}+1\Bigr)^{3-\mu}+\Bigl(\frac{b}{q}\Bigr)(2-\mu) \Bigr\}\\
term2&=&\biggl\{n\Bigl(\frac{b}{q}\Bigr)^{6}+8n\Bigl(\frac{b}{q}\Bigr)^{5}+18n\Bigl(\frac{b}{q}\Bigr)^{4}+(17n+1)\Bigl(\frac{b}{q}\Bigr)^{3}+(7n+6)\Bigl(\frac{b}{q}\Bigr)^{2}+(n+5)\Bigl(\frac{b}{q}\Bigr)+1 \biggr\} \times \nonumber \\
&& \Bigl\{\Bigl(\frac{b}{q}+1\Bigr)^{2-\mu}(3-\mu)+\Bigl(\frac{b}{q}\Bigr)(2-\mu)^{2} \Bigr\}\\
term3&=&\biggl\{n\Bigl(\frac{b}{q}\Bigr)^{6}+4n\Bigl(\frac{b}{q}\Bigr)^{5}+10n\Bigl(\frac{b}{q}\Bigr)^{4}+2(5n+2)\Bigl(\frac{b}{q}\Bigr)^{3}+3(n+6)\Bigl(\frac{b}{q}\Bigr)^{2}+10\Bigl(\frac{b}{q}\Bigr)+1 \biggr\} \times \nonumber \\
&& \Bigl\{\Bigl(\frac{b}{q}+1\Bigr)^{3-\mu}+\Bigl(\frac{b}{q}\Bigr)(2-\mu) \Bigr\}
\end{eqnarray}

\begin{eqnarray}
\frac{<t_{\lambda}^{(7)}>}{<t_{0}^{(7)}>}&=&\frac{num}{den}
\end{eqnarray}
where
\begin{eqnarray}
num&=&\biggl\{7\Bigl(\frac{b}{q}\Bigr)^{6}+56\Bigl(\frac{b}{q}\Bigr)^{5}+126\Bigl(\frac{b}{q}\Bigr)^{4}+120\Bigl(\frac{b}{q}\Bigr)^{3}+55\Bigl(\frac{b}{q}\Bigr)^{2}+12\Bigl(\frac{b}{q}\Bigr)+1 \biggr\} \times \nonumber \\
&& \Bigl\{n\Bigl(\frac{b}{q}\Bigr)^{\lambda}+n (\lambda-1)^{2}\Bigl(\frac{b}{q}\Bigr)^{2}+n \lambda \Bigl(\frac{b}{q}\Bigr)+(\lambda+1) \Bigr\}\\
den&=&\biggl\{7\Bigl(\frac{b}{q}\Bigr)^{6}+28\Bigl(\frac{b}{q}\Bigr)^{5}+70\Bigl(\frac{b}{q}\Bigr)^{4}+74\Bigl(\frac{b}{q}\Bigr)^{3}+39\Bigl(\frac{b}{q}\Bigr)^{2}+10\Bigl(\frac{b}{q}\Bigr)+1 \biggr\} \times \nonumber \\
&& \Bigl\{n\Bigl(\frac{b}{q}\Bigr)^{\lambda}+n (\lambda-1)\Bigl(\frac{b}{q}\Bigr)^{2}+n \Bigl(\frac{b}{q}\Bigr)+1 \Bigr\}
\end{eqnarray}

Interestingly, $<t_{\mu}^{(n)}>/<t_{0}^{(n)}>$ exhibits a maximum at a value of $b/q$ that shifts 
with the variation of $n$. For example, for the 3-state model the maximum occurs at 
\begin{eqnarray}
\frac{b}{q}=\frac{-1 + \sqrt{n-1}}{n} = 0.138071
\end{eqnarray}
whereas for the 5-state model  the maximum shifts to  
\begin{eqnarray}
\frac{b}{q}=0.120274
\end{eqnarray}

\begin{figure}[h]
\begin{center}
\includegraphics[angle=0,width=0.7\columnwidth]{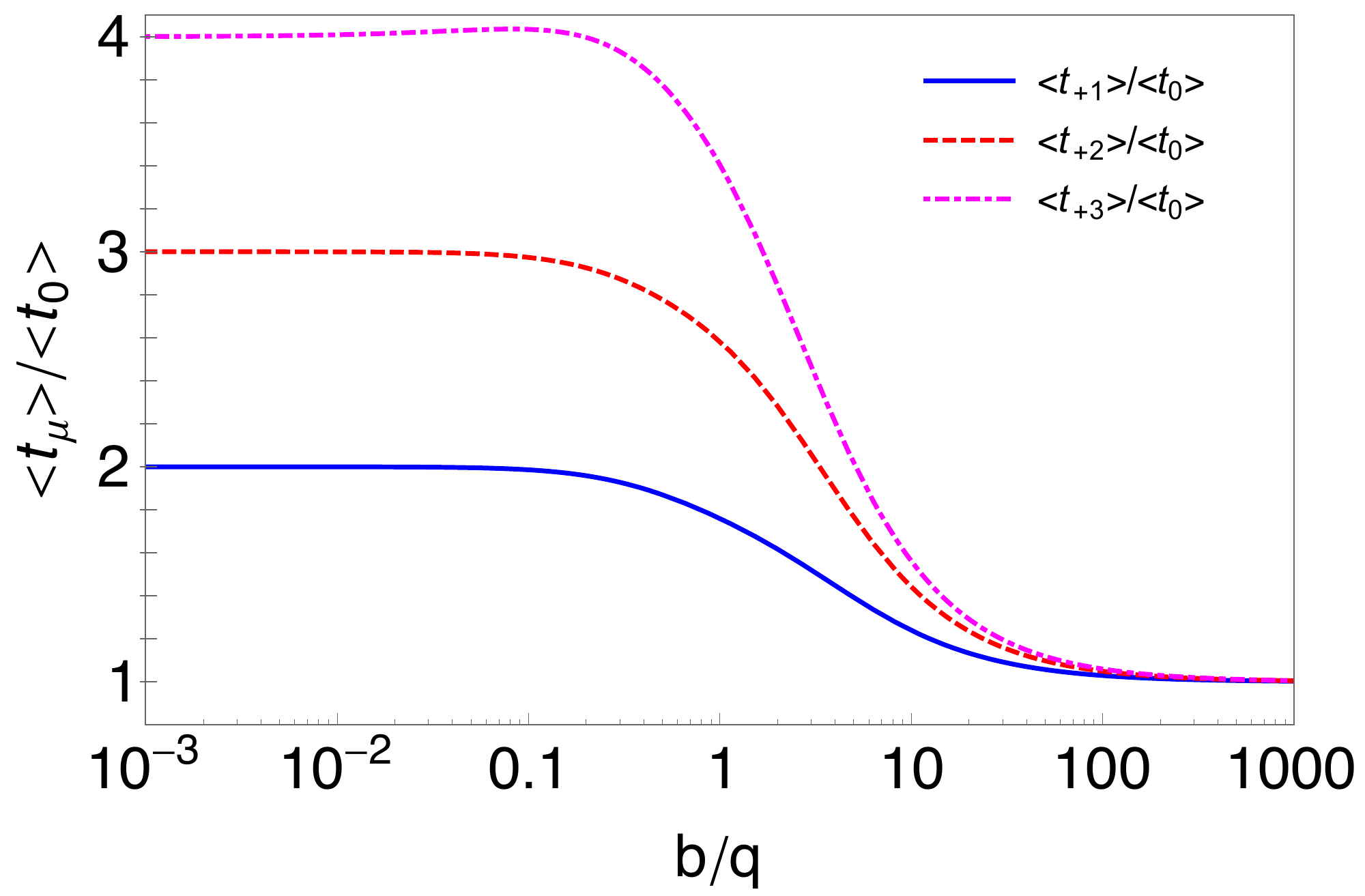}
\end{center}
\caption{$<t_{\mu}^{(n)}>/<t_{0}^{(n)}>$ are plotted against $b/q$ for the special value $n=7$ (i.e., 7-state model).}
\label{fig-7state}
\end{figure}


Note that the values of $<t_{\mu}^{(n)}>/<t_{0}^{(n)}>$ in the two limits 
$b/q \to  0$ and $b/q \to \infty$ reduce to very simple forms
\begin{eqnarray}
\lim_{\frac{b}{q}\to\ 0}  \frac{<t_{\mu}^{(n)}>}{<t_{0}^{(n)}>} &=& \mu+1 \\
\lim_{\frac{b}{q}\to\infty} \frac{<t_{\mu}^{(n)}>}{<t_{0}^{(n)}>} &=& 1.
\end{eqnarray}
which have clear physical meaning. In the limit $b/q \to \infty$ the population 
in all the distinct states corresponding to different $\mu$, for a given fixed $j$, get 
rapidly equilibrated (i.e., equally populated) and hence lead to 
$<t_{\mu}^{(n)}>/<t_{0}^{(n)}> = 1$ irrespective of $\mu$. In the opposite 
limit time taken to add each extra unit (i.e., each extra slippage of the nascent tape) 
adds up causing linear increase of  $<t_{\mu}^{(n)}>/<t_{0}^{(n)}>$ with $\mu$.

Finally, the numerical values of $b$ and $q$ for wild type RNAP under physiological conditions \cite{Pal03,Lin15,Olspert15,Olspert16,Parks14} 
are normally such that $b/q$ is of the order of $10^{-3}$ and hence $<t_{\mu}^{(n)}>/<t_{0}^{(n)}> \simeq \mu +1$. 
However, the rate of slippage, and hence the ratio $b/q$, can increase as much as almost tenfold in 
mutant RNAP \cite{Strathern13,Zhou13}. But, even tenfold increase of $b$ would not be adequate to 
observe a value of $<t_{\mu}^{(n)}>/<t_{0}^{(n)}>$ that deviates significantly from the limiting value 
$\mu+1$. Therefore, we suggest that, in addition to using the mutant RNAP for budding yeast 
({\it Saccharomyces Cerevisiae}), the RNAP should also be starved of nucleotides, monomeric subunits 
of the nascent transcript, so as to lower $q$ to a value comparable to $b$. Under such experimental 
conditions  the ratios $<t_{\mu}^{(n)}>/<t_{0}^{(n)}>$ can attain nontrivial values, different from $\mu+1$,  
that can be extracted from the exact expressions derived above.

\section{Summary and conclusion}

Fidelity of transcription by a RNA polymerase (RNAP) motor is achieved by the correct positining of its active site at  the tip of nascent mRNA. A backward or forward slippage of the nascent transcript with respect to the template DNA strand as well as RNAP motor results in a transcript that is, respectively, longer and shorter than the template. One of the fundamental questions is the time taken for successful transcription of one nucleotide on the template DNA by the RNAP during which the nascent RNA transcript associated with it becomes longer or shorter by $n$ nucleotides because of the net effects of backward and forward slippages. In our model we have treated the RNAP as a Tape-copying Turing Machine (TM) where the DNA template strand is the input tape while the nascent is the output tape. In the terminology of the TM the quantity of interest here is the time taken to complete a single successful computation during which the output tape becomes longer or shorter by $n$ units because of slippage. Identifying this time as a ``first-passage time'', we have derived exact analytical expression for the distribution of this first-passage time and hence the moments.

In two limiting cases the analytical expressions reduce to very simple forms which have interesting physical interpretations. In the intermediate regime we discover an interesting trend of variation that may have important implications in the biological context of transcript slippage which is one of the modes of {\it recoding} of genetic information \cite{atkins10}.

\section*{Appendix}
\section*{11-state model: symmetric case}
\begin{eqnarray}
P_{0}^{(11)} (s) &=& \frac{(b+q+s)^{5}+10b^{4}(q+s)+25b^{3}(q+s)^{2}+18b^{2}(q+s)^{3}+4b(q+s)^{4}}{(q+s)\{11b^{5}+55b^{4}(q+s)+77b^{3}(q+s)^{2}+44b^{2}(q+s)^{3}+11b(q+s)^{4}+(q+s)^{5}\}}\\ \nonumber
\\
P_{1}^{(11)} (s) &=& \frac{b(b+q+s)^{4}+6b^{4}(q+s)+9b^{3}(q+s)^{2}+3b^{2}(q+s)^{3}}{(q+s)\{11b^{5}+55b^{4}(q+s)+77b^{3}(q+s)^{2}+44b^{2}(q+s)^{3}+11b(q+s)^{4}+(q+s)^{5}\}}\\ \nonumber
\\
P_{2}^{(11)} (s) &=& \frac{b^{2}(b+q+s)^{3}+3b^{4}(q+s)+2b^{3}(q+s)^{2}}{(q+s)\{11b^{5}+55b^{4}(q+s)+77b^{3}(q+s)^{2}+44b^{2}(q+s)^{3}+11b(q+s)^{4}+(q+s)^{5}\}}\\ \nonumber
\\
P_{3}^{(11)} (s) &=& \frac{b^{3}(b+q+s)^{2}+b^{4}(q+s)}{(q+s)\{11b^{5}+55b^{4}(q+s)+77b^{3}(q+s)^{2}+44b^{2}(q+s)^{3}+11b(q+s)^{4}+(q+s)^{5}\}}\\ \nonumber
\\
P_{4}^{(11)} (s) &=& \frac{b^{4}(b+q+s)}{(q+s)\{11b^{5}+55b^{4}(q+s)+77b^{3}(q+s)^{2}+44b^{2}(q+s)^{3}+11b(q+s)^{4}+(q+s)^{5}\}}\\
P_{5}^{(11)} (s) &=& \frac{b^{5}}{(q+s)\{11b^{5}+55b^{4}(q+s)+77b^{3}(q+s)^{2}+44b^{2}(q+s)^{3}+11b(q+s)^{4}+(q+s)^{5}\}}
\end{eqnarray}

\section*{Acknowledgements}
This work is supported by ``Prof. S. Sampath Chair'' Professorship (DC) and a J.C. Bose National Fellowship (DC).

\end{document}